\documentclass[sigconf]{acmart}

\AtBeginDocument{%
  \providecommand\BibTeX{{%
    \normalfont B\kern-0.5em{\scshape i\kern-0.25em b}\kern-0.8em\TeX}}}

\setcopyright{acmcopyright}
\copyrightyear{2022}
\acmYear{2022}

\usepackage{algorithm}
\usepackage{indentfirst}
\usepackage{amsmath}
\usepackage{algpseudocode}
\usepackage{graphicx}

\begin{document}

\title{A case study of proactive auto-scaling for an ecommerce workload}

\author{Marcella Medeiros Siqueira Coutinho de Almeida}
\email{marcella.almeida@ccc.ufcg.edu.br}
\affiliation{%
  \institution{Federal University of Campina Grande}
  \city{Campina Grande}
  \state{Paraíba}
  \country{Brazil}
}

\author{Thiago Emmanuel Pereira}
\email{temmanuel@computacao.ufcg.edu.br}
\affiliation{%
  \institution{Federal University of Campina Grande}
  \city{Campina Grande}
  \state{Paraíba}
  \country{Brazil}
}

\author{Fabio Morais}
\email{fabio@computacao.ufcg.edu.br}
\affiliation{%
  \institution{Federal University of Campina Grande}
  \city{Campina Grande}
  \state{Paraíba}
  \country{Brazil}
}

\renewcommand{\shortauthors}{Almeida, Pereira and Morais}

\begin{abstract}
  Preliminary data obtained from a partnership between the Federal University of Campina Grande and an ecommerce company indicates that some applications have issues when dealing with variable demand. This happens because a delay in scaling resources leads to performance degradation and, in literature, is a matter usually treated by improving the auto-scaling. To better understand the current state-of-the-art on this subject, we re-evaluate an auto-scaling algorithm proposed in the literature, in the context of ecommerce, using a long-term real workload. Experimental results show that our proactive approach is able to achieve an accuracy of up to 94 percent and led the auto-scaling to a better performance than the reactive approach currently used by the ecommerce company.
\end{abstract}





\maketitle
\begin{sloppypar}

\section{Introduction}
Cloud computing refers to a model in which configurable computing resources (e.g., networks, servers, storage, applications, and services) are outsourced on an on-demand pay-per-use basis \cite{ilyushkin2018experimental}. Thus, the computing resources could be rapidly provisioned and released with minimal management effort or service provider interaction, providing an elastic infrastructure \cite{mell2011nist}. The auto-scaling techniques use this elasticity to automatically adjust resources based on the application demand running in the cloud infrastructure.

Specifically, auto-scaling can be defined as the process of dynamically adapting the resources assigned to the elastic applications, depending on the input workload, without the intervention of a human manager \cite{lorido2014review, morais2013autoflex, morais2017provisionamento}. In Amazon Web Services (AWS), this feature allows users to automatically launch or terminate virtual instances using its EC2 service based on predefined policies, schedules, and regular health status checks \cite{radhika2021review}. 

There are two ways of scaling any application: vertical scaling and horizontal scaling. The first one works on increasing (scale-up) and decreasing (scale-down) the number of computing resources assigned to each application instance, while the latter scales by increasing (scale-out) and decreasing (scale-in) the number of application instances \cite{rossi2019horizontal}. 

As for the classification of auto-scalers, when grouped by their anticipation capabilities \cite{lorido2014review}, it can be divided into two classes: reactive and proactive. In reactive, the user has to specify thresholds for workload metrics, meaning that scaling only occurs after such thresholds are exceeded. In proactive, an application takes early scaling decisions based on estimates of future demand \cite{shariffdeen2016adaptive}. 

Both strategies have trade-offs to take into account. While the reactive strategy is simple and intuitive \cite{radhika2021review} (as consequence is broadly utilized in commercial cloud providers \cite{lorido2014review}), the time taken by the auto-scaler to react and allocate resources could be too long to avoid performance degradation. On the other hand, the drawback of a proactive technique is that its reliability depends on the accuracy of the predicted values \cite{shariffdeen2016adaptive}.

The literature has shown a variety of approaches, evaluations, and comparisons between these two models and their own subcategories \cite{lorido2014review}. To the best of our knowledge, none of them has come to a conclusion about which one is better in terms of performance for ecommerce companies. In order to provide a better understanding of this matter, our goal is to evaluate the reactive approach already used by an ecommerce company compared to a proactive approach using the same configurations, simulating both auto-scaling behaviors with a real collected workload. For that, we implemented a time-series analysis technique, Autoregressive Integrated Moving Average.

Results show that our implemented model achieves an accuracy of 94\% and deviates very little from the actual data, less than 7\%. Even though a good prediction does not necessarily mean the model would make good allocation decisions, our performance results were pretty optimistic when using our forecast model in the auto-scaling simulation, since they were highly better than the ones from the reactive approach.
 
The rest of this work is structured as follows: Section 2 presents previous works on evaluating auto-scaling techniques, Section 3 exposes our prediction model, its implementation, and metrics for analysis, and Section 4 discusses our experimental results. Finally, Section 5 discloses our conclusions and possible future work.

\section{Related Work}

Auto-scaling and its techniques have been studied in many past works. Lorido-Botran et al. proposed a classification of these techniques into five main categories: static threshold-based rules, control theory, reinforcement learning, queuing theory, and time series analysis, using this classification for a literature review of previous proposals for auto-scaling \cite{lorido2014review}. More recently, Radhika and Sadasivam also conducted a comprehensive study on existing auto-scaling techniques, comparing newer proactive and reactive auto-scaling strategies but discussing possible research solutions for hybrid environments, too  \cite{radhika2021review}. Both these studies do not evaluate the suggested approaches.

On the other hand, Ilyushkin et al. presented a detailed comparative study of general state-of-the-art auto-scaling policies, this time conducting various experiments and comparing their performance in pairwise and group comparisons \cite{ilyushkin2018experimental}. Their work includes no predictive auto-scaling policies in their experiments and discussions, only reactive and hybrid techniques. 

As for studies comparing proactive auto-scaling approaches, Wang et al. proposed a predictive technique as a better alternative to the reactive one already used by Amazon EKS, with experiments using real-world workloads \cite{wang2021predicting}. Still, they deal only with vertical auto-scaling and a workload with a duration of one day. Islam et al. \cite{islam2012empirical} explored two predictive algorithms related to time-series analysis, incorporating sliding window to the training and prediction stages. Their benchmark is 135 minutes, ending with the same issue as Wang et al. \cite{wang2021predicting} of using a short trace.

Although Gao et al. used an extensive trace from Google Cluster Trace and evaluated different approaches of proactive auto-scaling (e.g., statistical, machine learning, and deep learning approaches), they did not deepen how ARIMA was implemented as the statistical approach \cite{gao2020machine}. Calheiros et al. implemented and compared ARIMA-based prediction to a trace of one week of an actual workload, but the workload of an ecommerce company was not included in their analysis \cite{calheiros2014workload}.

Considering the current state-of-art, most auto-scaling techniques have been evaluated using traces typically short spanning a few minutes, hours, or days \cite{papadopoulos2016peas}, and none of these evaluations were made testing a workload of ecommerce directly. With the crescent use of cloud computing and auto-scaling strategies for improvement, our work differs from the previous articles mentioned as it aims at analyzing which approach has a better performance for ecommerce traffic, conducting experiments with two weeks of data from an ecommerce company workload.

\section{Methodology}

The main goal of our work is to evaluate a proactive auto-scaling strategy compared to the reactive approach already used by an ecommerce company. To do so, we explored several existing predictive techniques through a bibliographical review, searching for which algorithm would fit better with the trace and metrics used by the ecommerce company from where the data were collected. We considered proactive models utilizing horizontal scaling and CPU load as the metric.

To better describe our methodology, the remainder of this section is divided as follows: Section 3.1 describes our dataset and how the data were collected. Sections 3.2 and 3.3 explains the reactive and the proactive strategies evaluated in this work, respectively, while Sections 3.4 and 3.5 detail how we implemented and simulated the auto-scaling behavior with our chosen techniques. Finally, Section 3.6 defines each metric used for our evaluation, of both prediction and performance.

\subsection{Data collected}

Our data were obtained from an application of the ecommerce company in which its current team believes that the application uses more virtual machines than needed at some moments of the day. We saw in this an opportunity to evaluate the impact of changing from a reactive auto-scaling strategy to a proactive one in case of significant infrastructure consumption. The data was collected in February 2022 from AWS Cloud Watch, with a trace of two weeks, from February 1st to 17th. The dataset we are working with has three important variables:

\begin{itemize}
    \item {\verb|timestamp|}: the moment of each observation, with a data grain of 60 seconds;
    \item {\verb|instance_type|}: the type of the instance, used to calculate the number of vCPUs;
    \item {\verb|utilization|}: the CPU utilization in that given moment.
\end{itemize}

This means that each minute we have new data to consider in our forecast, leaving us with historical data of more than 10 thousand observations to process before predicting. As for ``\verb|instance_type|'', the application we are analyzing uses Amazon EC2 instance types with four vCPUS available each \cite{ec2instances}, and its auto-scaling configurations will be discussed in the following section.

\subsection{Reactive strategy}

Widely studied in the past, reactive auto-scaling is usually based on threshold-based rules techniques and tends to deal with some intrinsic problems such as using cooldown times (also called inertia or calm) or dynamic thresholds \cite{ramperez2021flas}. The idea behind using the cooldown period is to prevent the auto-scaling from launching or terminating additional instances before the effects of previous activities are visible. The strategy of cooldown is not used for scheduled, periodic scaling actions or, in our case, predictive scaling.

In this study, we evaluate simple scaling, i.e., the current reactive strategy used by our chosen application. On Amazon EC2, this policy allows the user to create CloudWatch alarms for the scaling policies and specify the high and low thresholds for these alarms. A CloudWatch alarm is a feature that sends notifications when the chosen metrics fall outside of the levels (high or low thresholds) configured by the user. The main issue with simple scaling is, as mentioned before, the presence of cooldown periods: it must wait for the scaling activity or health check replacement to complete and the cooldown period to end before responding to additional alarms \cite{ec2policies}. 

In our application, the simple scaling policy was configured as follows:

\begin{itemize}
    \item Due to confiability matters, the exact capacity numbers will be omitted. So, it's important to know the min and max capacities were of x and 7.5x instances, respectively;
    \item The evaluation period was 5 min;
    \item The cooldown period was 6 min;
    \item The scaling rules were based on CPU utilization. The lower threshold was 25\%, and the upper one was 50\%. According to the rules:
    \begin{itemize}
        \item when CPU utilization was lower than 25\%, a number of y instances were removed;
        \item when CPU utilization was higher than 50\%, a number of 4y instances were added;
    \end{itemize}
\end{itemize}

\subsection{Proactive strategy}

As it is hard to dynamically identify and adjust optimal threshold values to maximize the utilization of the provisioned infrastructure, proactive auto-scaling tries to offer the solution to some of these challenges \cite{iqbal2018dynamic}. For our evaluation of one of the proposed solutions in the literature, we decided to take a better look at how a statistical approach would perform in comparison to a reactive one, without the higher complexity that a machine learning or deep learning technique would offer. Reviewing Calheiros et al. work, we noticed that their studied algorithm, ARIMA, would fit our mentioned parameters of horizontal scaling and CPU load, so we decided to reproduce this approach \cite{calheiros2014workload}.

Autoregressive Integrated Moving Average, i.e., ARIMA is a model utilized for time series prediction in different fields such as finance and economics \cite{box2015time}. It predicts a given time series based on its past values, which “autoregressive” stands for, and the goal is to reach stationary data that is not subject to seasonality \cite{duke}. According to Hyndman and Athanasopoulos \cite{hyndman2018forecasting}:

\begin{quote}
    If we combine differencing with autoregression and a moving average model, we obtain a non-seasonal ARIMA model. ARIMA is an acronym for AutoRegressive Integrated Moving Average (in this context, “integration” is the reverse of differencing). The full model can be written as \newline \newline
    $
    y'_{t} = c + \phi_{1}y'_{t-1} + \cdots + \phi_{p}y'_{t-p}
     + \theta_{1}\varepsilon_{t-1} + \cdots + \theta_{q}\varepsilon_{t-q} + \varepsilon_{t}
    $ \newline \newline
    where $y'_{t}$ is the differenced series (it may have been differenced more than once). The “predictors” on the right hand side include both lagged values of $y_t$ and lagged errors. We call this an $ARIMA(p,d,q)$ model, where\newline

    $p$ = order of the autoregressive part;
    
    $d$ = degree of first differencing involved;
    
    $q$ = order of the moving average part.
\end{quote}

For our implementation of the ARIMA model, we used the auto.arima() function in R. This function uses a variation of the Hyndman-Khandakar algorithm \cite{hyndman2008automatic}, which combines unit root tests, minimization of the AICc (Second-order Akaike Information Criterion), and MLE (Maximum Likelihood Estimation) to obtain an ARIMA model. This function is used to evaluate the model and estimate a configuration of the ARIMA parameters. 

\subsection{Algorithm implementation}

We used the R package forecast for the prediction, passing our collected historical data as a parameter in the auto.arima() function. In the context of this work, the historical data to be passed in our function means the observed number of CPU utilization in a determined timestamp, in our case, every 60 seconds, multiplied by the number of vCPUs available for each instance type, divided by 100. This calculation gives us the CPU demand in that given moment, and the function returns the predicted CPU demand in the horizon we settled as a parameter.

In this scenario, the horizon is based on the application start-up time, i.e., the time the application takes to initialize a VM. We need to predict with such an advance that by the time the application needs to add or remove instances, our algorithm has already decided on the correct action considering the time it would take for the start-up. The application chosen has a start-up time of between 9 and 12 minutes, so we opted for the worst-case scenario and settled our horizon to 12.

Another decision for our function was to incorporate the sliding window technique, meaning that it will consider new input values throughout the loop and leave older ones behind when predicting. The input $x$ is a vector of resource usage samples over $k$ consecutive time intervals, and the predicted output, $y$, is $r$ intervals ahead of this input window \cite{islam2012empirical, dietterich2002machine}. This left us with the following implementation:

\begin{algorithm}
\caption{Prediction model with ARIMA}\label{alg:cap}
\begin{algorithmic}[1]
\State \(D\) is a list with size N (\(N \in \mathbb{N}\)) representing CPU demand, in d should be the \(d^{th}\) CPU demand in list (\(d \in \mathbb{Q}^+\));
\State \(D[i, j]\) is a sub sequence of CPU demand, \(D[i,j] = \{d^{i}, d^{i + 1}, ..., d^{j - 1}, d^{j}\}\);
\State \(start\) and \(end\) are variables that define the range of the sliding window (\(start,end \in \mathbb{N}\));
\State \(ArimaForecast(L, H)\) is a function that tries to predict the next \(H\) elements of list \(L\), returns the last element \(p\) (\(H \in \mathbb{N}\), \(p \in \mathbb{Q}^+\), \(\exists_{i,j} D[i, j] = L\));
\For {each $i \in [initial\_value, final\_value]$}
\State \(prediction \leftarrow ArimaForecast(D[start,end], horizon)\);
\If{i + horizon <= N}
\State saves the last predicted value of \(prediction\) in a new list M;
\EndIf
\EndFor
\end{algorithmic}
\end{algorithm}

\subsection{Auto-scaling simulation}
To compare proactive and reactive auto-scaling techniques' performances, we need to simulate how each would behave with its respective policy configurations. For that, we used an open-source auto-scaling simulator, developed by the Laboratory of Distributed Systems \cite{simulator}, which already provides the simple scaling policy, and implemented our proactive scaling approach based on the function shown in the previous section.

This auto-scaling simulator receives a timestamp (the current time of the observation), the instance type, and the CPU utilization as input, returning the number of cores to be added or removed on the next observation based on the chosen policy configuration. The configuration for the proactive technique had a target value instead of upper and lower bounds like simple scaling, and it was calculated as an approximated value between 25\% and 50\% (the current configuration on our application), being settled as 35\%. It should be noted that this policy does not use cooldown periods when scaling.

Our model then had to predict the CPU demand for 12 minutes ahead and, to reach the target of 35\% in that future observation, it should decide to add or remove cores. It receives a training file of historical data and a prediction file so it gets new values as the sliding window goes by in the prediction. For our experiment, we've run with a week of data in our training file (from February 1st to 8th) and another week of data in our prediction file (from February 9th to 17th), analyzing the results with the chosen metrics reported in the next section.

\subsection{Evaluation metrics}

To evaluate our prediction's accuracy, we chose different statistical metrics: $R^2$ Prediction Accuracy, Mean Absolute Error (MAE), and Mean Absolute Percentage Error (MAPE)\cite{kotz2005encyclopedia, everitt2010cambridge}. These three metrics were selected because they are known for measuring accuracy in forecast models, showing different perspectives of how the prediction approximates the real data. 

Since a high accuracy of the forecast does not necessarily imply a good performance of our auto-scaling, we also need a metric to evaluate the results of our simulation. The performance evaluation for both strategies was made with the Auto-scaling Demand Index (ADI) metric \cite{netto2014evaluating}, used for analyzing and comparing system utilization. All these mentioned metrics will be described in the following subsections.

\subsubsection{$R^2$ Prediction Accuracy} 

The $R^2$ Prediction Accuracy measures how the prediction model approximates the real data points, with its value being within the range from 0 to 1, which means that the $R^2$ prediction accuracy of 1 indicates a perfect fit of the fitted model \cite{islam2012empirical}. The formula for this metric is:

\begin{equation}
    R^2=1-\frac{\sum_{i=1}^{n} (\hat{y_i}-{y_i})^2}{\sum_{i=1}^{n} (\hat{y_i}-\bar{y})}
\end{equation}

Where $\hat{y} = \frac{1}{n} \sum_{i=1}^{n} {y_i}$, ${y_i}$ is the actual output, $\hat{y_i}$ is the predicted output, and $n$ is the number of observations.

\subsubsection{Mean Absolute Error (MAE)} 

The Mean Absolute Error is a standard measure of forecast error in time series analysis and is the mean of the absolute difference between a target value and model prediction, i.e., the mean of all absolute errors. To reach this value, the formula is:

\begin{equation}
    MAE = \frac{1}{n} \sum_{i=1}^{n} \mid{\hat{y_i}} - {y_i}\mid
\end{equation}

Where $n$ is the number of errors and $\mid{\hat{y_i}} - {y_i}\mid$ the absolute errors. This calculation tells us an average of how big of an error it is expected from our prediction.

\subsubsection{Mean Absolute Percentage Error (MAPE)} 

The metric Mean Absolute Percentage Error measures the accuracy of a forecast model as a percentage and is given by the following formula:

\begin{equation}
    MAE = \frac{1}{n} \sum_{i=1}^{n} \frac{\mid{\hat{y_i}} - {y_i}\mid}{{y_i}}
\end{equation}

Where $n$ is the number of fitted points, ${y_i}$ is the actual value, and $\hat{y_i}$ is the forecast value. The lower the MAPE value, the better the prediction accuracy.

\subsubsection{Auto-scaling Demand Index (ADI)} 

Auto-scaling Demand Index is a performance metric for evaluating auto-scaling strategies. This metric is defined by the sum of all distances computed between each utilization level reported by the system and the target utilization interval, with upper and lower bounds \cite{netto2014evaluating}. The closer the result gets to zero, the better the auto-scaling performance, and it's calculated as follows:

\begin{equation}
    \sigma = \sum_{t \in \tau} 	{\sigma_t}
\end{equation}

where

\[
    {\sigma_t} = 
\begin{cases}
    L - {u_t},& \text{if } {u_t}	\leq L,\\
    0, & \text{if } L < {u_t} < U,\\
    {u_t} - U, & \text{otherwise.}
\end{cases}
\]

U and L are the values for the upper and lower bounds settled on the application configuration. Ut is the percentage of infrastructure utilization at a time-step T, and the following calculation gives it:

\[
    {u_t} = \frac{{w_t}}{{m_t}}
\]

Where ${w_t}$ is the number of resources demanded by the application (e.g., number of cores) and ${m_t}$ is the number of allocated resources (e.g., the sum of cores of the VMs running the application). 

Even though our proactive auto-scaling policy works with a single target, we calculated its ADI setting L and U values the same as the simple scaling configuration, 25\% and 50\%, respectively, since our goal is to have a system utilization between these thresholds and compare it with the performance of our reactive strategy.

\section{Results}
In our proactive approach, the auto-scaling simulator decides to add and remove resources based on the cores' demand predicted by ARIMA. To ensure it will make suitable decisions, we first need to check the accuracy of our model before analyzing our auto-scaling strategy results, so we evaluated the prediction accuracy using three of the metrics mentioned in the previous section, i.e., $R^2$ Prediction Accuracy, Mean Absolute Error (MAE), and Mean Absolute Percentage Error (MAPE).

 \begin{table}[htp]
   \caption{Results of each metric for our prediction model. They show a high accuracy since it deviates very little from the original values.}
   \label{tab:freq}
   \begin{tabular}{rrr}
     \toprule
     $R^2$ & MAE & MAPE (\%)\\
     \midrule
     0.94 & 2.50 & 6.82\\
   \bottomrule
 \end{tabular}
 \end{table}

\begin{figure}[htpb]
  \includegraphics[width=1\linewidth]{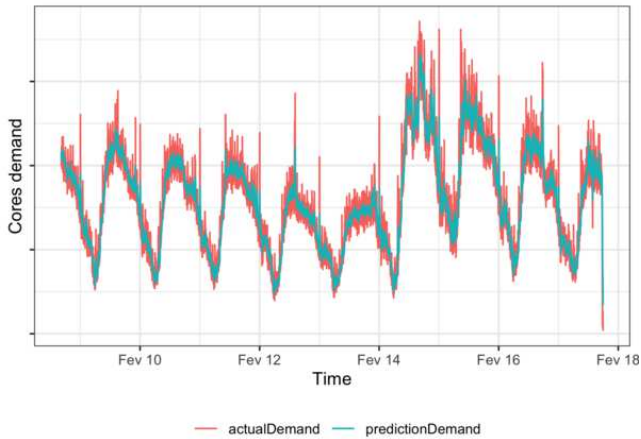}
  \caption{Actual cores' demand and predicted demand throughout February 9th and 17th. The similar behavior between them indicates a high accuracy of the forecast model, deviating an average of only 2,50 cores from the actual demand.}
  \label{fig:prediction}
\end{figure}

Figure \ref{fig:prediction} compares the collected data on cores’ demand and the prediction made. We can see very similar behavior, except for a few demand peaks in the actual data that our model could not predict. This similarity is supported by the numbers of MAE and MAPE, 2.50 and 6.82\%, respectively, meaning our forecast deviates less than 10\% from the actual demand and is very close to reality. The $R^2$ Prediction Accuracy had a result of 0.94 out of 1.0, which indicates our model is a good fit for explaining 94\% of the fitted data in the regression model. All these results are presented in Table \ref{tab:freq}.

\begin{figure}[htpb]
  \includegraphics[width=1\linewidth]{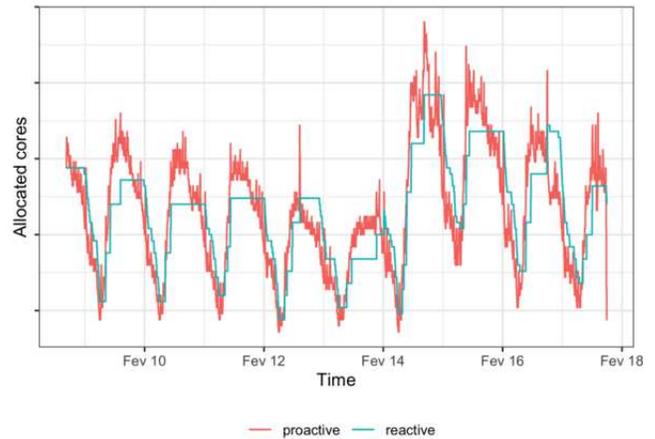}
  \caption{Allocated cores when using proactive and reactive strategies. The presence of cooldown periods in the reactive approach led it to a more conservative behavior when allocating cores and a variation of 30.19\% compared to a variation of 35.41\% from the proactive one.}
  \label{fig:allocated}
\end{figure}
\begin{figure}[htpb]
  \includegraphics[width=1\linewidth]{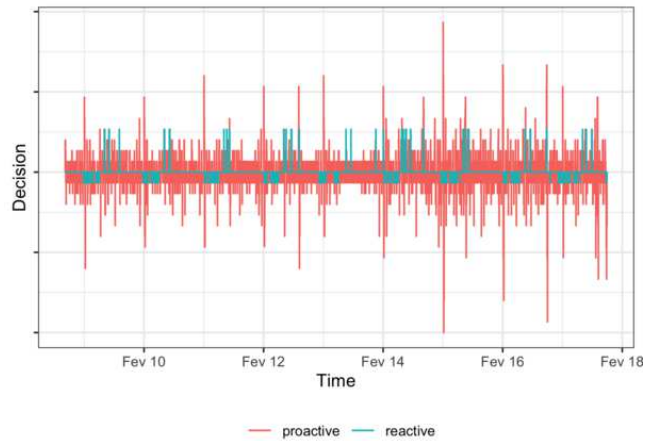}
  \caption{Adding and removing decisions from proactive and reactive strategies. The proactive approach takes way more decisions of both adding and removing cores, while the reactive one allocates way lower resources and spends long periods of time without making any decisions.}
  \label{fig:decision}
\end{figure}

We then run our auto-scaling simulator for both proactive and reactive strategies, noticing a higher variation in the decision to allocate cores when running the proactive technique and a conservative behavior from the reactive technique, as shown in Figure \ref{fig:allocated}. The coefficient of variation for proactive was 35.41\%, while for reactive was 30.19\%. This variation is also seen in Figure \ref{fig:decision}, where it shows the variable Decision, which indicates how many cores the algorithm decided to add or remove in each observation. 

\begin{figure}[htpb]
  \includegraphics[width=1\linewidth]{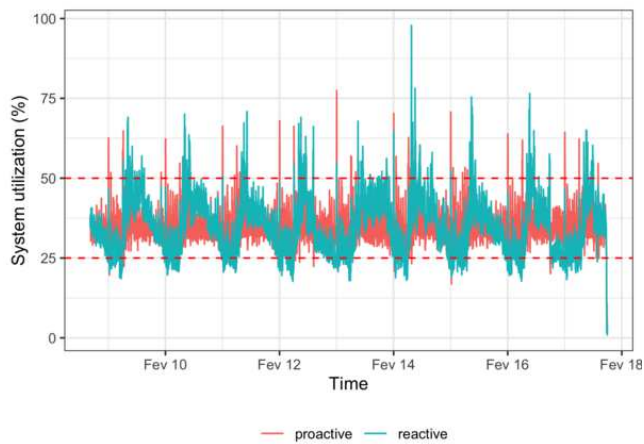}
  \caption{System utilization when running proactive and reactive strategies. With the proactive technique, the utilization tends to stay more between the upper and lower bounds, illustrated in the plot as the dashed red lines.}
  \label{fig:utilization}
\end{figure}

For the evaluation of each approach’s performance, we used the Auto-scaling Demand Index (ADI). This metric is calculated by taking values of system utilization and the settled thresholds, shown in Figure \ref{fig:utilization}. When calculating ADI, the proactive strategy got a result of 1782.97, while the reactive strategy got 7014.97. The goal is to get the result as close to zero as possible, considering it is the sum of the differences between the thresholds and the observed system utilization. This means our proactive approach reached a way better value than the reactive approach.

A conservative behavior such as in simple scaling may have led to a degradation of the performance, considering the time it takes to decide on adding or removing instances, as seen in Figure \ref{fig:decision}, where the reactive auto-scaling has extended periods of the variable Decision as zero. The anticipation of scaling up and down resources, and taking into account the start-up time for each VM in the application, could have been decisive in the divergence of performance results for proactive and reactive auto-scaling techniques.

\section{Conclusion}

In this work, we aimed to evaluate the performance of a proactive strategy compared to the performance of an already used reactive strategy by an ecommerce company. Based on our metrics, ARIMA appeared to be a good method to be considered for predictive technique implementation. An accuracy of 94\% was an optimistic first step in our experiments, especially considering that it was a forecast of more than 10 thousand observations ahead, with a horizon of 12 minutes for each prediction. 

Our auto-scaling simulation also showed a promising perspective when running proactive policy configurations. It was not guaranteed that a great forecast model would lead to great decisions of adding and removing instances, while also meeting our defined thresholds, but the results indicated, in fact, much better performance in reducing system utilization levels compared to the reactive approach. 

It should be noted, however, that this current work does not consider if a performance improvement would lead to a cost reduction for the ecommerce company since it did not analyze purchase plans and prices of VMs. For future studies, we plan to implement and evaluate other proactive approaches, incorporating a bigger trace of the same company workload and considering capacity planning in our analysis. 

\begin{acks}
This work wouldn’t be possible without the help and support of many people. An acknowledgment isn’t enough to my professors, Thiago Emmanuel and Fábio, who were the best mentors a student could ask for — all their patience and guidance were essential to finishing this work. To Ítallo and Fireman, who also helped me through this whole process. To each of my friends, who rooted for this achievement with me as if it were theirs. To my parents, Helda and Marcelo. They taught me the importance and the power of education in life, but the most important lesson I got from them is how love and comprehension give you the strength to pursue your dreams. I am pursuing mine thanks to them. To my grandfather, José Walter, who always wished to see his granddaughter graduating. I hope he can see it somehow.

This work has been funded by MCTIC/CNPq-FAPESQ/PB (EDITAL Nº 010/2021) and by VTEX BRASIL (EMBRAPII PCEE-1911.0140).
\end{acks}

\bibliographystyle{ACM-Reference-Format}
\bibliography{sample-base}

\end{sloppypar}
\end{document}